\documentclass{emulateapj}
\usepackage{amsmath}
\usepackage{graphicx}
\usepackage{amsfonts}
\usepackage{natbib}
\usepackage{color}
\usepackage{epstopdf}
\usepackage{graphicx}
\usepackage{multirow}

\begin{document}

\title[]{A Method for Locating High Energy Dissipation Region in Blazar}
\author{Dahai Yan\altaffilmark{1,2}, Qingwen Wu\altaffilmark{3}, Xuliang Fan\altaffilmark{3}, Jiancheng Wang\altaffilmark{1,2}, Li Zhang\altaffilmark{4}}
\altaffiltext{1}{Key Laboratory for the Structure and Evolution of Celestial Objects, Yunnan Observatory, Chinese Academy of Sciences, Kunming 650011, China; yandahai@ynao.ac.cn}
\altaffiltext{2}{Center for Astronomical Mega-Science, Chinese Academy of Sciences, 20A Datun Road, Chaoyang District, Beijing 100012, China}
\altaffiltext{3}{School of Physics, Huazhong University of Science and Technology, Wuhan 430074, China}
\altaffiltext{4}{Key Laboratory of Astroparticle Physics of Yunnan Province, Yunnan University, Kunming 650091, China}
\begin{abstract}

The production site of gamma-rays in blazar jet is an unresolved problem.
We present a method to locate gamma-ray emission region in the framework of one-zone emission model.
From measurements of core-shift effect, the relation between the magnetic
field strengths ($B'$) in the radio cores of jet and the distances ($R$) of these radio cores from central supermassive black hole (SMBH) can be inferred.
Therefore once the magnetic field strength in gamma-ray emission region ($B'_{\rm diss}$) is obtained, one can use the relation of $B'$-$R$ to derive the distance ($R_{\rm diss}$) of gamma-ray emission region from SMBH.
Here we evaluate the lower limit of $B'_{\rm diss}$ by using the criteria that the optical variability timescale $t_{\rm var}$ should be longer or equal to the synchrotron radiation
cooling timescale of the electrons that emit optical photons.
We test the method with the observations of PSK 1510-089 and BL Lacertae,
and derive $R_{\rm diss}<0.15\delta_{\rm D}^{1/3}(1+A)^{2/3}\ $pc for PSK 1510-089 with $t_{\rm var}\sim$ a few hours,
and $R_{\rm diss}<0.003\delta_{\rm D}^{1/3}(1+A)^{2/3}\ $pc for BL Lacertae with $t_{\rm var}\sim$ a few minutes.
Here $\delta_{\rm D} $ is the Doppler factor and $A$ is the Compton dominance (i.e., the ratio of the Compton to the synchrotron peak luminosities).
%The constraints for $R_{\rm diss}$ and $B'_{\rm diss}$ are independent of multi-wavelength spectra modeling.

\end{abstract}

\bigskip
\bigskip

%%%%%
%Section 1 - Introduction
%%%%%

\section{INTRODUCTION}
\label{sec:intro}
Blazars are one class of radio-loud active galactic nuclei (AGNs), pointing their relativistic jets at us.
According to the features of optical emission lines, blazars are usually divided into two classes: BL Lac objects (BL Lacs) with weak or even no observed optical emission lines and flat spectrum radio quasars (FSRQs) with strong optical emission lines
\citep{urry}.
Multi-wavelength radiations spanning from radio, optical to TeV gamma-ray energies have been observed from blazars.
Blazar emission is generally dominated by non-thermal radiation from relativistic jet.
The broadband spectral energy distribution (SED) of a
blazar has two distinct humps.
It is generally believed that the first hump is
the synchrotron radiation of relativistic electrons in the jet, however the origin of the
$\gamma$-ray hump is uncertain.

Leptonic and hadronic models
have been proposed to produce the second bump \citep[see, e.g.,][for review]{bhk12}.
In leptonic models, $\gamma$ rays are produced through inverse Compton (IC) scattering of high energy electrons,
including synchrotron-self Compton scattering  \citep[i.e., SSC; e.g.,][]{Maraschi1992,Tavecchio98,Finke08,Yan14} and external Compton (EC) scattering \citep[e.g.,][]{Dermer93,Sikora1994,bl,Dermer09,Paliya}.
$\gamma$ rays in hadronic models can be attributed to the processes including
proton- or pion-synchrotron radiation \citep{mann92,Aharonian2000,mucke2003} and $p\gamma$
interactions induced cascade \citep{bottcher09,bottcher13,murase14,masti13,Weidinger2015,Yan15,2015MNRAS.448..910C}.
In general,  both leptonic and hadronic models are able to reproduce the SEDs well,
but they require quite different jet properties. For instance the hadronic models require extremely high jet powers for the most powerful blazars, FSRQs \citep{bottcher13,Zdziarski}.

In blazar jet physics, an open question is the location of gamma-ray emission region, which controls the radiative cooling processes in both leptonic and hadronic models.
The location also means the place where the bulk
energy of the jet is converted to an energy distribution of high energy particles.
Because the gamma-ray emission region is usually compact, it cannot be directly resolved by current detectors. Many methods have been proposed to constrain the location of the gamma-ray emission region \citep[e.g.,][]{liu06,Tavecchio2010,Yan12,Dotson,Nalewajko14,Jorstad01,Jorstad10,Agudo,bottcher16}.
One popular method is to model the SEDs of FSRQs, and the location of the gamma-ray emission region (i.e., the distance from central back hole to the gamma-ray emission region) is treated as a model parameter \citep[e.g.,][]{Yan15b}.

Recently some methods independent of SED modeling are proposed to locate the gamma-ray emission region.
\citet{Dotson} suggested that the energy dependence of the decay times in flare profiles 
%in different-energy gamma-ray light curves 
could reflect the property of IC scattering.
If the decay times depend on gamma-ray energies, it indicates that IC scattering happens in the Thomson region where the electron cooling time due to IC scattering depends on the energy of the electron. This situation will occur when the gamma-ray emission region locates in dust torus where the seed photons for IC scattering have the mean energy of $\sim0.1\ \rm eV$ \citep{Dotson, Dotson15,Yan16} .
Moreover, the variability timescales of gamma-ray emissions also provide hints for the location of the gamma-ray emission region.
For instance fast gamma-ray variability indicates that the emission region is very compact, which is usually thought to be close to the central black hole \citep[e.g.,][]{Tavecchio2010}.

Currently there is
no consensus on the location of the high energy dissipation region.
The results given by the methods mentioned above are very inconsistent, from 0.01 pc to tens of pc \citep{Nalewajko14}.

As suggested by \cite{Wu}, we also use the relation of $B'$-$R$ derived in the measurements of radio
core-shift effect \citep[e.g.,][]{Os,Sokolovsky,zam}, to constrain the location of high energy dissipation in blazars.
In \citet{Wu}, the magnetic field strength in gamma-ray emission region was derived in modeling SED. Here we use optical variability timescale to constrain the magnetic field strength in gamma-ray emission region, and therefore our method is fully independent of SED modeling.

\section{Method and results}
\subsection{Method}
The variabilities of synchrotron radiations (e.g., variability timescale and time delay between emissions in different bands ) have been suggested to
estimate  comoving magnetic field \citep[e.g.,][]{Takahashi,bottcher03}.

Optical emission with fast variability from blazar is believed to be synchrotron radiation of relativistic electrons.
If electron cooling is dominated by synchrotron cooling, 
the cooling timescale of electron in comoving frame is given by~\citep[e.g.,][]{Tavecchio98}
\begin{equation}
 t'_{\rm cooling} = \frac{3}{4} \frac{m_ec^2}{\sigma_T c}(\gamma u_B)^{-1}
          = \frac{6\pi m_ec}{\sigma_T\gamma B'^2_{\rm diss}}\ ,
\label{tcool}
\end{equation}
where $u_B = B'^2_{\rm diss}/8\pi$ is the energy density of the magnetic field in comoving frame, $m_e$ is the mass of electron, $\sigma_T$ is the cross section of Thomson scattering, $\gamma$ is the electron energy.  Meanwhile, the observational synchrotron frequency is written as
\begin{equation}
 \nu_{\rm syn} = \frac{4}{3} \nu_L \gamma^2 \frac{\delta_{\rm D}}{1+z}
     \approx3.7\times10^6 \gamma^2 \frac{B'_{\rm diss}}{1\ \rm G} \frac{\delta_{\rm D}}{1+z}\ \rm Hz\  ,
\label{nu}
\end{equation}
where $\nu_L = 2.8\times10^6 (B'_{\rm diss}/1\ \rm G)\ \rm s^{-1}$ is the Larmor frequency, and $z$ is redshift.

The observational variability timescale  $t_{\rm var}$ can be taken as the upper limit for the  cooling timescale in observer frame $t_{\rm cooling}=t'_{\rm cooling}(1+z)/\delta_{\rm D}$, i.e., $t_{\rm var} \geq t_{\rm cooling}$. Then we can get the lower limit for magnetic field strength from equation~(\ref{tcool}) and~(\ref{nu}), i.e.,
\begin{equation}
 B'_{\rm diss} \geq 1.3\times 10^8 t_{\rm var}^{-2/3} \nu_{\rm syn}^{-1/3} \delta_{\rm D}^{-1/3} (1+z)^{1/3}\ \rm G\ ,
\label{estimatemag}
\end{equation}
where $t_{\rm var}$ is in unit of second and $\nu_{\rm syn}$ in unit of Hz.

If electron cooling is dominated by EC cooling in the Thomson
regime, the cooling timescale of the electron in equation~(\ref{tcool}) should be modified by a factor of $(1+k)^{-1}$ \citep{bottcher03}, and then the lower limit of $ B'_{\rm diss}$ in equation~(\ref{estimatemag}) is modified by a factor of $(1+k)^{-2/3}$. Here $k$ is the ratio of the energy densities between an external
photon field and the magnetic field in the comoving frame. $k$ can be replaced with Compton dominance $A$, i.e., the ratio of IC to synchrotron peak luminosity \citep[][]{Finke13}.

In analogy to the calculation of $A$ in \citet{Nalewajko17}\footnote{They defined Compton dominance as the ratio of Fermi-LAT luminosity above one GeV to the {\it WISE} luminosity at $3.6\ \mu$m.},  $A$ can be calculated as $A=L_{\rm 1-100\ \rm GeV}/L_{\rm optical}$,
where $L_{\rm 1-100\ \rm GeV}$ is the luminosity between 1 GeV to 100 GeV and $L_{\rm optical}$ is the optical luminosity. 

Radio telescopes have the capability to resolve the structure of
blazar jet on $\sim$pc scale.
Very long baseline interferometry (VLBI) observations showed that
core-shift effect (the frequency-dependent
position of the VLBI cores) is common in AGNs.
The core-shift effect is caused by synchrotron self absorption \citep[e.g.,][]{Blandford}.
Under the condition of the equipartition between the jet
particle and magnetic field energy densities, core-shift effect can be used to evaluate the magnetic field strength along the jet,
and a relation between the magnetic field strength ($B'$, in units of Gauss) and the distance along the jet ($R$, in unit of pc) was found,
i,e, \textbf{$B'\propto (R/1\ \rm pc)^{-1}\ \rm G$} \citep[e.g.,][]{Os,zam}.

Assuming that this relation still holds on in the sub-pc scale of the jet,
we then can use it and the lower limit for $B'_{\rm diss}$ derived by using optical variability to constrain the distance of gamma-ray emission region from SMBH ($R_{\rm diss}$).

\subsection{Results: testing the method with PKS 1510-089 and BL Lacertae}

Here we use observations of two blazars to test the feasibility of our method.

PKS 1510-089 ($z$=0.361) is a TeV FSRQ. Using four
bright gamma-ray flares detected by \emph{Fermi}-LAT in 2009, \cite{Dotson15} located its high energy dissipation region in dust torus (DT).
In May 2016, H.E.S.S. and MAGIC detected a very high energy (VHE) flare from PKS 1510-089 \citep{1510}. During the VHE flare, its optical emission also presented activity.
The R band (the frequency is $4.5\times10^{14}\ $Hz) flux decreased from $1.4\times10^{-11}\ \rm erg\ cm^{-2}\ s^{-1}$ to $1.1\times10^{-11}\ \rm erg\ cm^{-2}\ s^{-1}$ within $\sim$ 2 hr \citep[see Fig. 2 in][]{1510}.
We adopt the variability timescale $t_{\rm var}\approx2\ $hr.

In hadronic models, the electron cooling in a FSRQ is dominated by synchrotron cooling \citep[e.g.,][]{bottcher13,Diltz}.
Using equation~(\ref{estimatemag}) we derive  $B'_{\rm diss} \geq5\delta_{\rm D}^{-1/3}\ $G.
With this magnetic field strength, we can constrain the distance $R_{\rm diss}$ using the relation of $B'\approx 0.73\cdot (R/1\ \rm pc)^{-1}\ $G provided by \citet{zam},
and derive $R_{\rm diss}<0.15\delta_{\rm D}^{1/3}\ $pc.

In leptonic models, the electron cooling in a FSRQ is dominated by EC cooling. Then we have $B'_{\rm diss} \geq5\delta_{\rm D}^{-1/3}(1+A)^{-2/3}\ $G,
and $R_{\rm diss}<0.15\delta_{\rm D}^{1/3}(1+A)^{2/3}\ $pc.

For {\it Fermi}-LAT FSRQs,  $A$ is in the range from 0.1 to 30 \citep{Finke13,Nalewajko17}.
\citet{Dermer14} showed that for 3C 279 the value of $k$ varies from $\sim$3 to 20 from low states to high states.
For PKS 1510-089, \citet{Saito} derived $k\sim20$ from the SED during a $\gamma$-ray flare in March 2009.

Taking $\delta_{\rm D}=30$ and $A\sim k=20$, we derive $R_{\rm diss}<0.5\ $pc for hadronic models, and $R_{\rm diss}<3.5\ $pc for leptonic models.
Note that $A$ is sensitive to observing time for FSRQs, hence an $A$ obtained from simultaneous observation should be used in practice.

The sizes of broad line region (BLR) and dust torus can be estimated with the disk luminosity $L_{\rm disk}$ \citep{ghisellini09,ghisellini14}:
$$r_{\rm BLR}=10^{17}(L_{\rm disk}/10^{45}\rm \ erg\ s^{-1})^{1/2}\ \rm cm\ ,$$
$$r_{\rm DT}=10^{18}(L_{\rm disk}/10^{45}\rm \ erg\ s^{-1})^{1/2}\ \rm cm\ .$$
For PKS 1510-089, $L_{\rm disk}\approx5.9\times10^{45}\ \rm erg\ s^{-1}$ \citep{cas},
we obtain $r_{\rm BLR}\approx0.1\ $pc and $r_{\rm DT}\approx0.8\ $pc.

Because BLR photons will attenuate gamma-ray photons above $\sim30/(1+z)\ $GeV\footnote{Assuming that the BLR radiation field is dominated by $\rm Ly\alpha$ line photons with the mean energy of $\approx10\ $eV.}, the detection of VHE photons from PKS 1510-089 indicates that its gamma-ray emission region should be outside the BLR.
Our results support the scenario that the gamma-rays of PKS 1510-089 are produced in dust torus.
%\citet{Dai} reported that on 2000 May 29 the R band magnitude of PKS 1510-089 faded from $3.4\times10^{-12}\ \rm erg\ cm^{-2}\ s^{-1}$ to $7.0\times10^{-13}\ \rm erg\ cm^{-2}\ s^{-1}$ in 0.5 hr. This short variability timescale leads to  $B'_{\rm diss} \geq5.3\ $G, and $R_{\rm diss}<0.14\ $pc.
This is consistent with the result in \citet{Dotson15}.

This method is also used to constrain the location of gamma-ray emission region in BL Lacs.
\citet{Os} derived a relation of $B'\approx 0.14\cdot (R/1\ \rm pc)^{-1}\ $G for BL Lacertae (2200+420; $z$=0.069).
\citet{Covino} reported a very fast optical variability for this source. On 2012 September 1 the R-band flux decayed by a factor of about 3 in $5\ $min.
This fast variability requires $B'_{\rm diss} \geq31.3\delta_{\rm D}^{-1/3}\ $G for hadronic models, and $B'_{\rm diss}\geq31.3\delta_{\rm D}^{-1/3}(1+A)^{-2/3}\ $G for leptonic models. Then we have $R_{\rm diss}<0.004\delta_{\rm D}^{1/3}\ $pc for hadronic models and $R_{\rm diss}<0.004\delta_{\rm D}^{1/3}(1+A)^{2/3}\ $pc for leptonic models.

For {\it Fermi}-LAT BL Lacs, $A$ is in the range from 0.1 to 3 \citep{Finke13,Nalewajko17}.
\citet{Abdo} showed that from a low state to a flare state, $k$ varies from $\sim0.1$ to 3 for BL Lacertae.
Using $\delta_{\rm D}=30$ and $k=3$, we have $R_{\rm diss}<0.01\ $pc for hadronic models and $R_{\rm diss}<0.02\ $pc for leptonic models.

\citet{Raiteri} estimated that the accretion disk luminosity $L_{\rm disk}$ is $6\times10^{44}\rm \ erg\ s^{-1}$ for BL Lacertae. The energy density of the photon field attributed to accretion disk radiation at $R_{\rm diss}=0.02\ $pc is  $\sim$0.4$\ \rm erg\ cm^{-3}$ which is much greater than the energy density of BLR photon field of $\sim$0.01$\ \rm erg\ cm^{-3}$ \citep[e.g.,][]{ghisellini09,Hayashida}.
However this situation prohibits the production of VHE photons because of $\gamma$-$\gamma$ absorption by accretion disk photons and BLR photons.
Therefore the detection VHE photons \citep[e.g.,][]{Arlen}  cannot be accompanied with a fast optical variability with $t_{\rm var}\sim 5\ $min.

Here we just aim to present the constraints on $R_{\rm diss}$ given by using various optical variability timescales, i.e., the feasibility of our method.
From the above descriptions, one can find that our method is very effective, especially for the source having fast optical variability.
In a specific study it is better to choose simultaneous optical flare respect to gamma-ray emission to obtain variability timescales and Compton dominance,
and the definition of the variability timescale should be clarified.

\section{Discussion}

Our method for locating the gamma-ray emission region relays on two assumptions:

(1) optical and gamma-ray emissions are produced in the same region;

(2) the relation of $B'$-$R$ obtained from radio
core-shift measurements can be extrapolated into sub-pc scale of jet.

In general, the first assumption still works in the current blazar science,
although a class of orphan gamma-ray flares seems challenge the one-zone emission model \citep[e.g.,][]{MacDonald}.
For the second assumption,
\citet{Os} extended the relation to the distance of $10^{-5}\ $pc at the SMBH (very close to the black hole jet-launching
distance), and found that the extrapolated magnetic field strengths are in general consistent with that expected from theoretical
models of magnetically powered jets \citep[e.g.,][]{Blandford}. So far, the above assumptions are reliable.

Besides the two assumptions, our constraint slightly relies on the value of Doppler factor, $\propto\delta_{\rm D}^{1/3}$;
while it depends on Compton dominance $A$ in leptonic models, $\propto(1+A)^{2/3}$.
In addition to the relation of $B'$-$R$, our method only requires simultaneous $\gamma$-ray and optical observations. 

The stringency of our constraint mainly depends on the precision of the measurement for the relation of $B'$-$R$.
\citet{Pushkarev} and \citet{zam} derived this relation for over 100 blazars by measuring the core-shift effect.
Combining the measurement of this relation and optical variability timescale, one can independently constrain the location of gamma-ray emission region in blazar.

We use two TeV blazars, PSK 1510-089 and BL Lacertae, to test our method.
Using the R-band variability with $t_{\rm var}\approx2\ $hr for PKS 1510-089,
we derive $R_{\rm diss}<0.15\delta_{\rm D}^{1/3}(1+A)^{2/3}\ $pc for leptonic models and $R_{\rm diss}<0.15\delta_{\rm D}^{1/3}\ $pc for hadronic models.
Using a typical value $\delta_{\rm D}=30$ and a large enough value $A=20$, we derive $R_{\rm diss}<0.5\ $pc for hadronic models and $R_{\rm diss}<3.5\ $pc for leptonic models.
%the inferred magnetic field strength of $B'_{\rm diss} \geq1.6\ $G for PKS 1510-089 is higher than the values of $B'_{\rm diss} \leq1\ $G  that are typically derived in the SED modeling with the leptonic models \citep[e.g.,][]{bottcher13,Saito,Wu}.
%For such a magnetic field strength, the $B'\propto (R/1\ \rm pc)^{-1}\ \rm G$ relation requires emission region locating within 0.46 pc from the SMBH, i.e., $R_{\rm diss}<6\ r_{\rm BLR}/0.6\ r_{\rm DT}$. The result is basically in agreement with the constraints derived by \citet{Dotson15} and \citet{Nalewajko14}.
%It should be noted that our method does not require an assumption of the gamma-ray emission mechanism, which is needed in the methods of \citet{Dotson15} and \citet{Nalewajko14}.

For BL Lacertae,
%its location of high energy emission region is not effectively constrained before.
%Therefore the emission mechanism is less certain. One cannot distinguish whether its gamma-rays are produced by SSC or EC \citep[e.g.,][]{Abdo}.
%Identifying the location of gamma-ray emission region as well as independent estimate of $B'_{\rm diss}$ could be helpful for us to determine high-energy emission mechanism for BL Lacertae.
we use a very short optical variability timescale of $t_{\rm var}\approx5\ $min reported in \citet{Covino} to estimate the lower limit for magnetic field strength,
and derive $R_{\rm diss}<0.003\delta_{\rm D}^{1/3}\ $pc for hadronic models and $R_{\rm diss}<0.003\delta_{\rm D}^{1/3}(1+A)^{2/3}\ $pc for leptonic models.
Using a typical value $\delta_{\rm D}=30$ and a large enough value $A=3$, we have $R_{\rm diss}<0.01\ $pc for hadronic models and $R_{\rm diss}<0.02\ $pc for leptonic models.
%$B'_{\rm diss} \geq15.6\ $G. Such a high value is comparable to the values of $\sim$10 G derived in hadronic models \citep[e.g.,][]{bottcher13}.
%Its high energy emission region should  be within 0.009 pc from the central black hole for such a rapid optical variability.
%For an intra-day variability, assuming $t_{\rm var}\sim10\ $hr, one can derive $B'_{\rm diss} \geq0.65\ $G and $R_{\rm diss}<0.22\ $pc.

One can see that with various optical variability timescales from minutes to a few hours,
the high energy emission region can be located within pc or subpc scale from central black hole in the framework of one-zone emission model.
%The result only depends on the measurement of the relation $B'\propto R^{-1}\ \rm G$.
The lower limit for the distance of the emission region from SMBH can be estimated by the absorption of GeV-TeV photons from low energy photons around jet \citep[e.g.,][]{liu06,bai,bottcher16}.

By modeling blazar SED, one can determine emission mechanisms and physical properties of the relativistic jets \citep[e.g.,][]{ghisellini10,ghisellini14,Kang,zhang15}.
In the previous studies, $B'_{\rm diss}$, $R_{\rm diss}$, and other model parameters are fitted together.
There are degeneracies between model parameters \citep[see][for correlations between model parameters given by Markov Chain Monte Carlo fitting technique]{Yan13,Yan15b}.
Our method provide independent constraints for $B'_{\rm diss}$ and $R_{\rm diss}$, and break degeneracies between model parameters.
This will lead to better understandings of emission mechanisms and physical properties of the relativistic jets.

It should be noted that the relation of $B'$-$R$ is derived under the assumption of the equipartition
between electron and magnetic field energy densities \citep[e.g.,][]{Os}. On the aspect of SED modeling,
it is found that the SEDs of FSRQs and low-synchrotron-peaked BL Lacs (LBLs) can be successfully fitted at the condition of ({\em near}-)equipartition between
electron and magnetic field energy densities in leptonic models \citep[e.g.,][]{Abdo,Yan16b,Hu},
while the leptonic modeling results for the SEDs of high-synchrotron-peaked BL Lacs (HBLs) are far out of equipartition \citep[e.g.,][]{Dermer15,Zhu}.
Therefore, for consistency, our method is applicable for FSRQs and LBLs in the framework of leptonic models.
The jet equipartition condition in hadronic models is rather complex. The modeling results are inconsistent \citep[e.g.,][]{bottcher13,Diltz}.
%\cite{Diltz} modeled the SED of 3C 279 using a hadronic model, and find that the best modeling result is near-equipartition between the energy densities of magnetic field and particles. At least for FSRQs, it is possible that the ({\em near})-equipartition condition could be achieved in hadronic models.
It is unknown whether the ({\em near})-equipartition condition could be achieved in hadronic models.

\section{Conclusion}
We presented an effective method for constraining the location of gamma-ray emission region in blazar jet in the framework of one-zone emission model.
Our method uses the relation of $B'$-$R$ derived in the VLBI core-shit effect in blazar jet.
The lower limit for magnetic field strength in gamma-ray emission region is estimated by utilizing the fact that the optical variability timescale should be
longer or equal to the synchrotron radiation cooling timescale of the electrons that produce optical emission.
Then the upper limit for the location of gamma-ray emission region is derived with the relation of $B'$-$R$.
%Besides the relation of $B'\propto R^{-1}$, only ({\em quasi-})simultaneous (respect to the gamma-ray emission) optical observation is needed in our method.
Our method is applicable for LBLs and FSRQs.

$\\$
% Acknowledgements %

We thank the anonymous referee for helpful comments
which significantly improved the paper. We acknowledge the financial supports from the National
Natural Science Foundation of China (NSFC-11573026, NSFC-11573060, NSFC-1161161010, NSFC-11673060, NSFC-U1738124).
D. H. Yan is also supported by the CAS ``Light of West China" Program.

%%%%%%
% Bibliography %
%%%%%%

\bibliography{ApJ}

%%%%%%%%%%%%%
\clearpage
%%%%%%%%%%%%%%%% lk_1m %%%%%%%%%%%%
%%%%%%%%%%%%%%%%%%%%%%
\end{document}